\documentstyle[aps,multicol,epsfig]{revtex}
\begin{document}
\draft
\title{First-principles investigation of spin polarized conductance in atomic carbon wires}
\vspace{1cm}
\author{L. Senapati\footnote[1]{Present Address: Department of Chemistry and Pitzer Center for Theoretical Chemistry, \\  University of California, Berkeley, CA 94720-1460, USA, email: senapl@socrates.berkeley.edu}, R. Pati, M. Mailman and S. K. Nayak\footnote[2]{For Correspondence: senapl@socrates.berkeley.edu, nayaks@rpi.edu}}

\address{Department of Physics, Applied Physics and Astronomy,\\ Rensselaer Polytechnic Institute,\\ Troy, NY 12180, USA.}
\maketitle

\begin{abstract}
  
We analyze spin-dependent energetics and conductance for one dimensional (1D) atomic carbon wires consisting of terminal magnetic (Co) and interior nonmagnetic (C) atoms sandwiched between gold electrodes, obtained employing first-principles gradient corrected density functional theory and Landauer's formalism for conductance. Wires containing an even number of interior carbon atoms are found to be acetylenic with $\sigma-\pi$ bonding patterns, while cumulene structures are seen in wires containing odd number of interior carbon atoms, as a result of strong $\pi$-conjugation. Ground states of carbon wires containing up to 13 C atoms are found to have anti-parallel spin configurations of the two terminal Co atoms, while the 14 C wire has a parallel Co spin configuration in the ground state. The stability of the anti-ferromagnetic state in the wires is ascribed to a super-exchange effect. For the cumulenic wires this effect is constant for all wire lengths. For the acetylenic wires, the super-exchange effect diminishes as the wire length increases, going to zero for the atomic wire containing 14 carbon atoms.   Conductance calculations at the zero bias limit show spin-valve behavior, with the parallel Co spin configuration state giving higher conductance than the corresponding anti-parallel state, and a non-monotonic variation of conductance with the length of the wires for both spin configurations.
\end{abstract}

\begin{multicols}{2} 
\section{Introduction}
   Miniaturization from sub-micron conventional solid state devices to extreme small scale single organic molecule based devices has been the focus of intensive research in recent years, motivated by the emerging field of molecular scale electronics and quantum information technology. Controlled transport of electrons in molecular wires containing only a few atoms forms the basis of molecular scale electronics. Significant recent advances in experimental techniques have made it possible to fabricate nano-wires containing only a few atoms and measure their electrical properties.$^1$ Specifically, atomic carbon wires containing up to 20 atoms have been synthesized.$^2$ These carbon wires serve as ideal models to develop understanding of and to eventually control the mechanism of electron transport in finite one-dimensional (1D) systems.  Previous theoretical and experimental studies$^{3-8}$ on atomic and molecular wires have been primarily limited to charge transport, with some recent exceptions.$^{9-14}$ Recent experimental measurements have shown that electron spin polarization can persist considerably longer than charge polarization.$^{15}$ It is consequently highly desirable to learn how to manipulate and enhance the control of electron transport offered by spin degrees of freedom, adding another dimension to the emerging field of molecular scale electronics and revealing important information for potential applications in spin-based molecular electronics (spintronics) as well as in quantum information processing.

   Pure carbon clusters have a long history.$^{16,17}$ Clusters with less than 10 atoms are known to have low-energy linear structures characterized by cumulenic bonding (C=C=C=C) with near equal bond lengths. These structures are stabilized by strong $\pi$-conjugation between the double bonds which are alternately directed in the x- and y-planes perpendicular to the bonds.$^{16}$ Some of these clusters also possess cyclic forms, which also become the stable form for larger sizes.$^{17}$ Lang and Avouris calculated the conductance of such a cumulenic carbon atom wire, i.e., with all C-C bond lengths constrained to be equal, connected on both ends directly to metal electrodes.$^5$ They found an oscillatory behavior, with wires composed of odd numbers of carbon atoms having a higher conductance than even-numbered wires. This was contrary to expectations based on a simple molecular orbital theory of the cumulene structure, and was attributed to electron donation from the metal contacts into additional $\pi$-bonds formed between the terminal carbon atoms and the electrodes.

   The presence of terminal magnetic atoms has been recently been shown$^{14}$ to modify both the structures and conductance properties of these wires. Pati et al have reported first-principles calculations$^{14}$ of spin-dependent electronic structures and energetics, as well as spin-polarized conductance of small carbon wires containing up to five carbon atoms that are terminated by magnetic atoms which are in term attached to gold electrodes. The magnetic atoms can act as spin polarizers or filters, resulting in a strong spin valve effect. These results showed that when terminated by magnetic atoms, the $\pi$-conjugated structure is not necessarily the lowest energy structure. In particular, for the even-numbered carbon wires, the acetylenic structure with alternating $\sigma$- and multiple $\pi$-bonds becomes more stable. The calculations also showed that the ground states of these magnetically terminated carbon wires have anti-parallel terminal atom spin configurations, which could be rationalized by contribution from super-exchange effects. The conductance of the wires was seen to non-monotonous with wire length just as in the pure carbon wires,$^5$ and is shown to depend strongly on the magnetic configuration of the terminal atoms, with the result that the wires could be made to act as a molecular spin valve.

   This first study of spin-dependent properties of short carbon wires$^{14}$ raised a number of fundamental questions. In particular, what happens when the length of the carbon wire spacer between the magnetic atom species is increased? How do the wire structures and the relative energetics of the different spin configurations change? how do the ground state spin configurations depend upon the length?  How does the effective exchange coupling between the magnetic species change with the number of carbon atoms in the wire? What is the critical length of the connecting wire up to which the super-exchange effects identified for the short wires survive?  

In order to address these questions and to thereby improve our understanding of spin-dependent electron transfer in extended molecular systems, we extend here the work in ref.14 up to wires containing 14 carbon atoms. Our first principles calculations, which explicitly include spin-polarization effects, reveal that the anti parallel Co spin state remains the ground state for wires with up to 13 carbon atoms, and also show a continuation of the alternation between cumulenic and acetylenic structures for odd and even wire lengths respectively. However, the 14 carbon atom wire is seen to have a parallel Co spin configuration in the ground state. Interestingly, we find that for wires containing an odd number of carbon atoms the energy difference (${\Delta} E$) between the anti-ferromagnetic and ferromagnetic state (anti-ferromagnetic being the ground state) remains constant as a function of  wire length. In contrast, in the acetylenic carbon chains, this energy difference decreases exponentially as a function of the number of atoms with the exception of the 2-carbon atom wire.  Analysis of this change in ground state spin configuration (14 carbon atom wire) in terms of the super-exchange contribution allows us to estimate the characteristic length for super-exchange in acetylenic carbon wires as $\sim$20~\AA.   We also find that the $\pi$-conjugated cumulenic wires exhibit higher conductance than the acetylenic wires. Finally,  the calculated magneto-conductance for different wire lengths shows a large difference between the two magnetization states, particularly for C-wires containing 13 and 14 C-atoms, suggesting its potential applications in  molecular magneto-electronics.

The remainder of the paper is organized as follows. Our computational approach is described in Section II.  The results and discussions are presented  in Section III.  Section IV summarizes our  results.   
  
\section {Computational Details}
 As in the previous study of short atomic wires,$^{14}$ we utilize an architecture consisting of chains of non magnetic C-atoms connecting two magnetic Co atoms. The Co-(C)$_n$-Co wire structures, with n=6 to 14, are subsequently inserted between two metal gold electrodes for calculation of spin polarized conductance. In a magnetic system like this, the total conductance can be evaluated as:     
\begin {equation}
g_t=g_{spin-conserved} + g_{spin-flip} , 
\end {equation}
\noindent where g$_{spin-conserved}$  is the conductance from the spin conserved part and g$_{spin-flip}$ is the conductance due to spin flip scattering. The latter contribution plays a significant role only when the spin-orbit coupling plays a significant role. Since the spin flip scatterthe spin-orbit coupling effect in highly ordered, strongly conjugated C-wire structures is expected to be small, leading to a large spin-flip scattering length, we have assumed the scattering to be coherent and have not included relativistic spin-orbit coupling effects in the present paper. The spin conserved part of the conductance is calculated as:
\begin {equation}
g_{spin-conserved}=g^\alpha + g^\beta, 
\end {equation}
\noindent where g$^\alpha$ and g$^\beta$ are the contribution to conductance from up (a) and down (b) spin states, respectively. Since at low bias the conduction primarily occurs in the close proximity of the Fermi energy of the metal contact, we can use Landauer's approach$^{18}$ to calculate and at the Fermi energy. In the zero bias limit, we have:
\begin {equation}
g^{\alpha(\beta)}{(E_f)} = \;\frac{e^2}{h}\;\ {T^{\alpha(\beta)}(E_f)}, 
\end {equation}   
\noindent where ${T^{\alpha(\beta)}(E_f)}$ is the transmission function for the spin up (${\alpha}$) or spin down (${\beta}$) electrons. This is evaluated using the Green's function derived from the Kohn-Sham matrix obtained from self-consistent spin unrestricted Density Functional calculations.$^{19}$  We have employed a gradient corrected Perdew-Wang 91 exchange and correlation functional$^{19}$ and double numerical polarized basis set$^{20}$ for the calculation of energetics and magnetic structures. Both the spin configurations and geometry for parallel and anti-parallel magnetic states between the Co atoms are simultaneously optimized using the self-consistent DFT approach. Anti-parallel magnetic configurations between the Co atoms are obtained by making use of the broken symmetry formalism. Details of this procedure can be found in refs ${13}$ and ${14}$. From the calculated spin-polarized conductance, we then estimate the magneto conductance (MC) according to:                  
\begin {equation}
MC = \;\frac{g_t(\uparrow,\uparrow)-g_t(\uparrow,\downarrow)}{g_t(\uparrow,\uparrow)},
\end {equation} 
 \noindent where ${g_t(\uparrow,\uparrow)}$ and ${g_t(\uparrow,\downarrow)}$ are given by the total conductance, Eq.(1) in the parallel and anti-parallel configurations, respectively.
 
\section{Results and Discussion} 
   \subsection{ Structures, magnetic properties, and energetics}

Using the procedures summarized in the previous section, we have optimized both the spin state and geometry for the Co-(C)$_{n=6-14}$-Co wire structures in parallel and anti parallel magnetic configurations between the Co atoms. Similar to the earlier report for short atomic wires containing up to 5 carbon atoms.$^{14}$, we find a clear $\pi-\pi$ and ${\sigma-\pi}$ bonding pattern for the C-atoms in the wire. This is illustrated in Fig. 1 for n=11 and 12 carbon atom wires, which shows the ground state 
\begin{figure}[ht]
\epsfig{figure=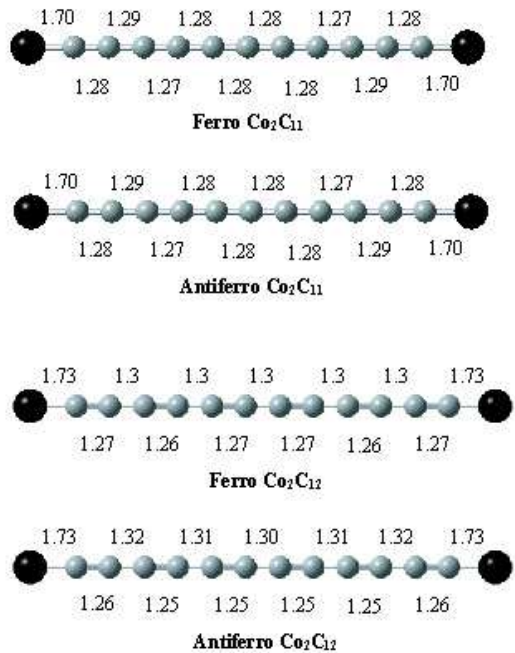,width=210pt}
\caption{Schematic of the bonding pattern in a carbon wire containing (a) 11 and (b) 12- carbon atoms. Terminal black spheres are Cobalt atoms and gray spheres are carbons. All the optimized bond distances are indicated for both parallel(ferro) and anti-parallel (antiferro) spin configurations of the atomic carbon wire.}
\end{figure}
\noindent structures of both magnetic configurations. The bond distances in the even carbon wires show a clear alternation, both for ferro and antiferro, consistent with ${\sigma-\pi}$ bonding. antiferro, consistent with ${\sigma-\pi}$ bonding. In contrast, the odd wire show no evidence of bond alternation, consistent with pure $\pi$-conjugated structure. Comparing the energy for parallel(ferro) and anti parallel(antiferro) spin configurations in the wires, we find that the anti-ferromagnetic state is lower in energy for all wires studied here, except n=14, which shows a ferromagnetic ground state. The calculated energy difference between ferromagnetic and anti ferromagnetic configurations, ${\Delta}E=E(\uparrow,\downarrow)-E(\uparrow,\uparrow)$, is shown as a function of number of C-atoms in the atomic wire in Fig. 2. For comparison purposes we have also shown here the results for short atomic wires obtained in $Ref. 14$. The energy difference between the two magnetization states are found to be larger than $k_BT$ at room temperature, suggesting that these antiferro-magnetic states are stable in normal operating conditions. The lower energy for the anti ferromagnetic spin configuration between the terminal Co atoms can be attributed to a super-exchange interaction that is facilitated by strong overlap of the magnetic Co and the non-magnetic C-atoms. A careful analysis of ${\Delta}E$ 
\begin{figure}[ht]
\epsfig{figure=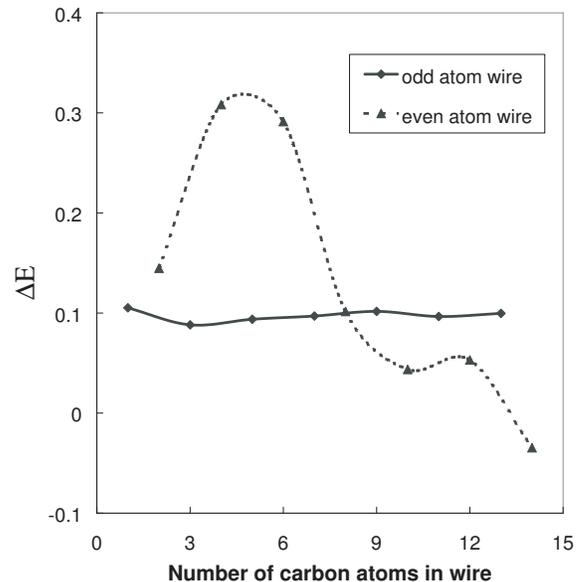,width=220pt}
\caption{The energy difference (${\Delta}E=E(\uparrow,\downarrow)-E(\uparrow,\uparrow)$) between parallel and anti parallel spin configurations in carbon atom wires, as a function of the number of carbon atoms in the wire, n. Even n chains are connected with a dashed line and odd n chains are connected with a solid line.  Values for n=1-5 carbon atoms are taken from $Ref. 14$. The energy difference(${\Delta}E$) is given in eV.
}
\end{figure}
\noindent
as function of wire length suggests that for the strongly $\pi$-conjugated wires i.e. cumulenic structures (odd number of C-atoms), the energy difference is approximately independent of the number of C-atoms in the wire. For $\sigma-\pi$ conjugated C-wires (even number of C-atoms),  ${\Delta}E$ exhibits an non-monotonic behavior with wire length. In particular, the anti-ferromagnetic state for C-wires containing 2, 4 and 6 carbon atoms are more stable than those for the $\pi$-conjugated C-wires.  For acetylenic wire, ${\Delta} E$ decreases in an exponential manner (with the exception of the 2 carbon atom wire) and is found to be negative for the wire containing 14 C-atoms. This suggests that the super exchange, which stabilizes the anti-ferromagnetic phase in $\sigma-\pi$ conjugated system, attenuates exponentially and becomes negligible for a wire containing 14 C-atoms. This allows us to estimate the super-exchange characteristic length for a $\sigma-\pi$ conjugated system to be $\sim$20~\AA. 

The additional stability for the wires containing 2, 4 and 6 C-atoms compared to 1, 3 and 5 C-atom wires could be explained  as follows.  In wires containing 1, 3 and 5 C-atoms, the exchange interaction is facilitated between the two terminal Co atoms through delocalized  spins shared by both the Co atoms stabilizing further the ferromagnetic coupling compared to that in short even atom wire. Similar ferromagnetic stabilization has been seen in the Fe$^{3+}$-Fe$^{2+}$ compounds.$^{21}$ This extra stability of ferromagnetic ordering in the short odd atom wire leads to smaller ${\Delta}E$ as compared to wires containing 2, 4 and 6 C-atoms.   Thus both double exchange and super exchange play an important role in stabilizing the magnetic ordering in these systems. In odd C-wires,  double exchange and super exchange effects remain constant due to continuous p-conjugation.  However, the super exchange effect exceeds the double exchange effect resulting in  anti ferromagnetic configurations as the ground states. In even C-wires containing 2,4 and 6 C-atoms, the double exchange effect is smaller compared to 1, 3, and 5 C-wires,  respectively. This is due to the lack of spin delocalization (Fig. 3) in even wires, which in effect destabilizes ferromagnetic ordering leading to a  large energy difference  between the two magnetic configurations (${\Delta}E$ in Fig. 2).  
\begin{figure}[ht]
\epsfig{figure=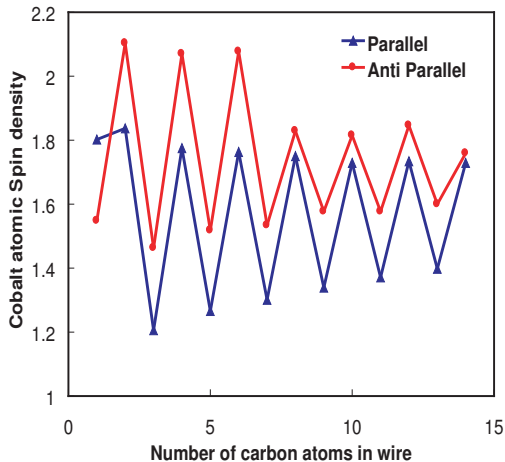,width=210pt}
\caption{Atomic spin density of the terminal Co atoms for both parallel and anti-parallel Co spin configurations, shown as a function of the number of carbon atoms in the wire, n.
}
\end{figure}  
\subsection{ Spin Polarized Conductance.}
Using Landauer's approach as outlined in Section II, we have calculated the spin-polarized conductance in the zero bias limit. The results are summarized in Fig. 4.  Several interesting features are apparent here. First, the conductance in the C-wire is found to be higher for parallel than for anti-parallel spin configuration of the terminal Co-atoms. This is a prerequisite for the spin-valve effect, which is primarily due to spin dependent scattering and which has been observed in magnetic/non-magnetic hetero bulk structures.$^{22,23}$ Second, both parallel and anti-parallel spin configurations show oscillations in conductance as a function of the wire length. For the parallel spin configurations, the conductance oscillation is damped after n=8 C-atoms and remains almost constant at about $1g_0$ $(g_0={2e^2/h})$ for the wire with n=12, 13 and 14 C-atoms. In contrast, in the anti-parallel case, the conductance is seen to decrease as n increases and to finally vanish for a wire for 12 and 14 C-atoms. The faster decrease of conductance with the wire length for even C-wires in the anti parallel case is due to the presence of $\sigma$-bonds in these systems, which can act as tunnel barriers for electron conduction. In fact, recent calculations on  $\sigma$-bonded structures  have  shown that the tunnel barrier increases with the increase of wire length$^{24}$, leading to an exponential decay in the electronic conduction. In contrast, in odd C-atom wires, the $\pi$-orbitals are highly delocalized, providing pathways for electron transfer and consequently leading to higher conductance as seen in Fig. 4. As discussed above, the super-exchange effect vanishes for a wire containing 14 C-atoms.  
 \begin{figure}[ht]
\epsfig{figure=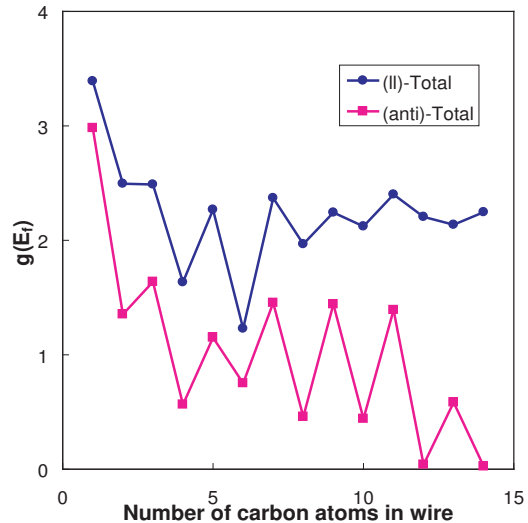,width=210pt}
\caption{Conductance (in units of ${e^2/h}$) evaluated at the Fermi energy for both parallel (ferro) and anti parallel (antiferro) spin configurations in the carbon wires, as a function of the number of carbon atoms in the wires, n.
}
\end{figure}
 
To understand the oscillatory pattern in conductance, we have also calculated the Mulliken charges and spin densities at individual atoms for different wire lengths.  The spin density at the Co atoms, i.e. the difference between the number of spin up and spin down electrons, are shown in Fig. 3. We see that for both parallel and anti-parallel magnetization states, the spin density at the Co-atoms oscillates with the number of C-atoms in the wire. Also, the $\sigma-\pi$ conjugated wires show a higher atomic spin density at Co than the $\pi$-conjugated wires.  This is not surprising since, as noted above, the $\pi$-conjugated systems have a stronger delocalization of spin compared to that in the $\sigma-\pi$ conjugated wires. We have also calculated the magneto conductance $(MC)$ according to Eq.(4). Fig.5 summarizes the ${MC}$ values as a function 
\begin{figure}[ht]
\epsfig{figure=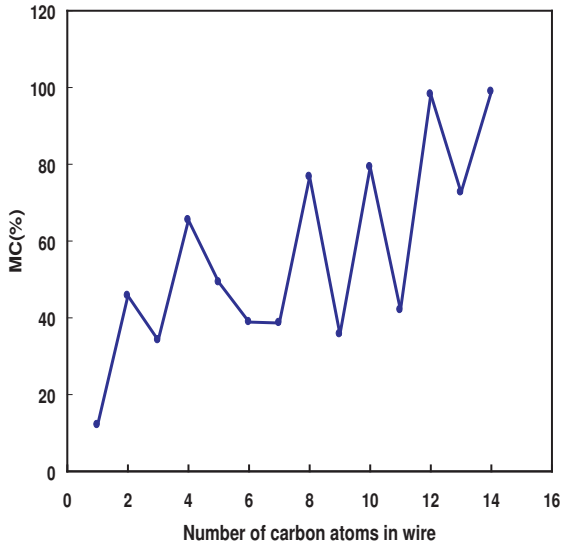,width=220pt}
\caption{Dependence of magneto conductance, MC, eq.(4), on the number of carbon atoms in the atomic wire, n.
}
\end{figure}
\noindent of number of C-atoms in the wires, with ${MC}$ displayed as a percentage. We find an oscillatory behavior in the magneto conductance, with a maximum value of $100\%$ change in magneto resistance between the parallel and anti parallel magnetization states for wires containing 12 and 14 C-atoms. This huge change in resistance between the two magnetization states suggests potential useful applications of these nanoscale materials for molecular magneto-electronics.

\section{Conclusions}

We have investigated the chain length dependent magnetic structures and energetics associated with highly conjugated C-wires sandwiched between magnetic Co atom species using the gradient corrected density functional approach. The Co-terminated wires show an alternation of structure between cumulenic for odd numbers of C atoms and acetylenic for even numbers of C atoms.  The spin-polarized conductance was calculated as a function of number of C-atoms in the wire in the zero bias limit using Landauer's formalism.  These length dependent calculations reveal an oscillatory pattern in conductance, with a significantly higher conductance arising for the parallel magnetization state compared to that for the anti-parallel magnetization state. The ground state C-wire structures containing up to 13 carbon atoms are found to have an anti-ferro magnetic spin configuration of the terminal Co atoms.  In contrast, the wire with 14 carbon atoms is found to have a parallel Co spin configuration in the ground state. The energy difference between the parallel and anti parallel  magnetization states is found to be larger than $k_BT$  at room temperature, suggesting that these two magnetization states are not interchangeable at normal operational temperatures. The stability of the anti ferromagnetic spin configuration between the terminal Co atoms is seen to arise because of a super-exchange interaction that is facilitated by strong orbital overlap of the terminal magnetic Co atoms and the non-magnetic C-atoms of the wire. This effect vanishes for the C-wire containing 14 C-atoms explaining the switch to a more stable parallel spin configuration. For carbon wires containing 12 and 14 C-atoms, we found almost no conductance in anti-parallel spin configurations, suggesting that the characteristic length for super-exchange interaction in these $\sigma-\pi$ conjugated carbon wires is about 20~\AA. A maximum value of $100\%$ change in magneto resistance was obtained for carbon wires containing 12 and 14 carbon atoms.

{\bf Acknowledgments}
We thank K.B. Whaley and J. Schrier for useful discussions on the length dependence of bond alternation and super-exchange, and for helpful comments on the manuscript. SKN also would like to thank Professor Z. Soos for helpful discussions. This work was supported by the NSF funded Nanoscale Science and Engineering Center at RPI. This work was also partially supported by National Computational Science Alliance under Grant Nos. MCA01S014N and DMR020003N, and by the ACS Petroleum Research Fund.

\end{multicols}
\end{document}